\documentclass[a4paper,10pt,twoside]{cpc-hepnp}

\usepackage{multicol}
\usepackage{graphicx}
\usepackage{epstopdf}
\usepackage{booktabs}
\usepackage{amssymb,bm,mathrsfs,bbm,amscd}
\usepackage[tbtags]{amsmath}
\usepackage{lastpage}

\def\p{\partial}
\def\s{\sigma}

\def\G{\Gamma}
\def\g{\gamma}

\def\d{\delta}
\def\de{\delta}

\def\ld{\lambda}
\def\Ld{\Lambda}

\def\th{\theta}
\def\e{\eta}

\def\om{\omega}
\def\Om{\Omega}
\def\rh{\rho}

\def\b{\beta}

\def\a{\alpha}

\def\pdellx'{\frac{\partial}{\partial x'}}
\def\pdellw'{\frac{\partial}{\partial w'}}

\newcommand{\be}{\begin{equation}}
\newcommand{\ee}{\end{equation}}
\def\bed{\begin{displaymath}}
\def\eed{\end{displaymath}}
\def\bea{\begin{eqnarray}}
\def\eea{\end{eqncrray}}
\def\[{$$}
\def\]{$$}

\begin{document}

\fancyhead[c]{\small Chinese Physics C~~~Vol. xx, No. x (201x) xxxxxx}
\fancyfoot[C]{\small 010201-\thepage}

\footnotetext[0]{Received 16 July 2018}


\title{Exact Recession Velocity and Cosmic Redshift  Based on\\ Cosmological Principle and Yang-Mills Gravity}
\author{%
Jong-Ping Hsu$^{1;1)}$\email{jhsu@umassd.edu} and Leonardo Hsu$^{2}$%
}
\maketitle

\address{%
$^1$ Department of Physics, University of Massachusetts Dartmouth, North Dartmouth, MA 02747-2300, USA\\
}

\address{%
$^2$ College of Education and Human Development, 
University of Minnesota, Minneapolis, MN 55455, USA\\
Department of Chemistry and Physics,  Santa Rosa Junior College,
Santa Rosa, CA 95401, USA} 

\begin{abstract}
  Based on the cosmological principle and quantum Yang-Mills gravity in the super-macroscopic limit, we obtain an exact recession velocity and cosmic redshift z, as measured in an inertial frame $F\equiv F(t,x,y,z).$   For a matter-dominated universe, we have the effective cosmic metric tensor 
$G_{\mu\nu}(t)=(B^2(t),-A^2(t),-A^2(t),-A^2(t)),  \ A\propto B\propto t^{1/2}$, where $t$ has the operational meaning of time in $F$ frame.   We assume a cosmic action $S\equiv S_{cos}$ involving $G_{\mu\nu}(t)$ and derive the `Okubo equation' of motion, $G^{\mu\nu}(t)\p_\mu S \p_\nu S - m^2=0$, for a distant galaxy with mass $m$.  This cosmic equation predicts an exact recession velocity, $\dot{r}=rH/[1/2 +\sqrt{1/4+r^2H^2/C_o^2} ]<C_o$, where $H=\dot{A}(t)/A(t)$ and $C_o=B/A$, as observed in the inertial frame $F$.  For small velocities, we have the usual Hubble's law $\dot{r} \approx rH$ for recession velocities.  Following the formulation of the accelerated Wu-Doppler effect, we investigate cosmic redshifts z  as measured in $F$. It is natural to assume the massless Okubo equation, $G^{\mu\nu}(t)\p_\mu \psi_e \p_\nu \psi_e=0$, for light emitted from accelerated distant galaxies.  Based on the principle of limiting continuation of physical laws, we obtain a transformation for covariant wave 4-vectors between and inertial and an accelerated frame, and predict  a relationship for the exact recession velocity and cosmic redshift, $z=[(1+V_r)/(1-V_r^2)^{1/2}] - 1$, where $V_r=\dot{r}/C_o<1$, as observed in the inertial frame $F$. These predictions of the cosmic model are consistent with experiments for small velocities and should be further tested.
\end{abstract}

\begin{keyword}
Yang-Mills gravity; recession velocity; speed limit of recession velocity;  redshift.
\end{keyword}

\begin{pacs}
11.15.-q,  98.80.-k 
\end{pacs}
\footnotetext[0]{\hspace*{-3mm}\raisebox{0.3ex}{$\scriptstyle\copyright$}2019
Chinese Physical Society and the Institute of High Energy Physics
of the Chinese Academy of Sciences and the Institute
of Modern Physics of the Chinese Academy of Sciences and IOP Publishing Ltd}%

\begin{multicols}{2}
\section{Introduction}

In previous work, Yang-Mills gravity with the principle of gauge symmetry (i.e., the spacetime translational ($T_4$) gauge symmetry) was formulated in general reference frames (both inertial and non-inertial) based on flat spacetime.  In the geometric-optics limit, the wave equations of quantum particles in Yang-Mills gravity lead to Hamilton-Jacobi type equations of motion with an effective metric tensor $G_{\mu\nu}(x)$ for classical objects and light rays.   We have shown that classical Yang-Mills gravity is consistent with experiments such as the perihelion shift of Mercury, deflection of light by the sun, redshifts and gravitational quadrupole radiation.\cite{1,2,3}  Yang-Mills gravity has also been quantized in inertial frames and gravitational Feynman-Dyson rules and the S-matrix have been obtained.\cite{2,4}  This theory has brought gravity back to the arena of gauge field theory and quantum mechanics.  It has also provided a solution to difficulties in physics such as the lack of an operational meaning of space and time coordinates\cite{5} in the conventional model of cosmology based on general relativity and the incompatibility between `Einstein's principle of general coordinate invariance and all the modern schemes for  quantum mechanical description of nature.'\cite{6}  This `most glaring incompatibility of concepts in contemporary physics' was discussed in details by Dyson in the 1972 Josiah Willard Gibbs Lecture, given under the auspices of the American Mathematical Society.

For readers not familiar with Yang-Mills gravity, we will first briefly summarize the key ideas and main equations.   Quantum Yang-Mills gravity is based on an external spacetime translational group ($T_4$) and involves (Lorentz) vector gauge functions $\Ld^\mu(x)$ in flat space-time with inertial frames.\cite{1,2,3}  As a result, the difficult problem of the quantization of the gravitational field, as discussed by Dyson, disappears.  
The $T_4$ gauge fields are massless spin-2 symmetric tensor fields $\phi_{\mu\nu}$. They are associated with  the $T_4$ group and the generator $p_{\mu}=i \p_{\mu}$ in inertial frames. The $T_4$ gauge covariant derivatives are obtained through the following replacement,\cite{1,2}
\be
\p_{\mu} \to \p_{\mu} - i g\phi_{\mu}^{\nu}p_{\nu} = J_{\mu}^{\nu}\p_{\nu},  \ \ \ \  \ \ 
J_{\mu}^{\nu}= (\d_{\mu}^{\nu} + g\phi_{\mu}^{\nu}),
\ee
in inertial frames with the metric tensor $\e_{\mu\nu}=(1,-1,-1,-1)$ and in natural units with $c=\hbar=1$. 

  As usual, the $T_4$ gauge curvature $C_{\mu\nu\a}$ is derived from the commutator of the gauge covariant derivative $J_{\mu}^{\ld}\p^\ld$, 
\be
[J_{\mu}^{\ld}\p_{\ld}, J_{\nu}^{\s}\p_{\s}]=C_{\mu\nu\a}\p^\a, 
\ee
\be
 C_{\mu\nu\a}=J_{\mu\ld}(\p^\ld J_{\nu\a})-J_{\nu\ld}(\p^\ld J_{\mu\a}), \ \ \ \  J_{\mu\ld}=J_\mu^\b \e_{\b\ld}.
\ee

The action $S_{\phi\psi}$ of Yang-Mills gravity for the tensor field $\phi_{\mu\nu}$ and a charged fermion field $\psi$ in an inertial frame is quadratic in the gauge curvature $C_{\mu\nu\a}$,
\be
S_{\phi\psi}=\int L_{\phi\psi} d^4 x, \ \ \ \   L_{\phi\psi}= L_{\phi} + L_{\psi},
\ee
\be   
L_{\phi}= \frac{1}{4g^2}\left (C_{\mu\nu\a}C^{\mu\nu\a}- 2C_{\mu\a}^{ \ \ \  \a}C^{\mu\b}_{ \ \ \  \b} \right),
\ee
\be
L_{\psi}=+ \overline{\psi}i\g^{\mu}(\p_\mu +g\phi_\mu^\nu \p_\nu -ie A_\mu) \psi  - m\overline{\psi}\psi.  
\ee
 The action $S_{\phi\psi}$ is invariant under local $T_4$ gauge transformations, although the Lagrangian density $L_{\phi}$ by itself is not invariant due to the presence of a total derivative term, which does not contribute to the gravitational field equations.\cite{3}  The structure of the Yang-Mills action in (4), (5) and (6) for the spin-2 tensor field $\phi_{\mu\nu}$ coupled to the charged fermion field $\psi$ are different from those in teleparallel gravity and Einstein-Cartan gravity.  These theories of gravity are not formulated in inertial frames and, hence, cannot be quantized to derive Feynman-Dyson rules for the calculation of the S matrix.\cite{3}
 
The wave equation of the gravitational tensor field $\phi_{\mu\nu}$ can be derived from the Lagrangians (5) and (6), $H^{\mu\nu}=g^2 S^{\mu\nu}$, where $H^{\mu\nu} \equiv \p_\ld (J^{\ld}_\rho C^{\rho\mu\nu} - J^\ld_\a C^{\a\b}_{ \ \ \ \b}\eta^{\mu\nu}) +.....$ and $S^{\mu\nu}$ is the source tensor of the fermion matter field.\cite{1,2}  It is interesting to observe that its linearized equation takes the form,
$$
\p_\ld \p^\ld \phi^{\mu\nu} -  \p^\mu \p_\ld \phi^{\ld\nu} -
\eta^{\mu\nu} \p_\ld \p^\ld \phi  + \eta^{\mu\nu} \p_\a \p_\b \phi^{\a\b}
$$
\be
+  \p^\mu \p^\nu \phi - \p^\nu \p_\ld \phi^{\ld\mu} - g S^{\mu\nu} = 0, \ \ \ \  \phi=\phi^\s_\s,
\ee 
which is formally the same as the linearized Einstein equation in general relativity.\cite{3}

 In quantum Yang-Mills gravity, the symmetric tensor field $\phi_{\mu\nu}$ is a massless spin-2 gauge boson.  The gravitational quadrupole radiation has been discussed with the usual gauge condition $\p^\mu \phi_{\mu\nu}=\p_\nu \phi^\ld_\ld /2$.  This gauge condition and the linearized equation (7) lead to the usual retarded potential, which can be expressed in terms of the polarization tensor $e_{\mu\nu}$.  For the symmetric polarization tensor of the massless tensor field in flat space-time, there are only two physical states with helicity $\pm 2$ that are invariant under the Lorentz transformations.
We have also calculated the power emitted per unit solid angle in the direction ${\bf x}/|{\bf x}|$ and that radiated by a body rotating around one of the principal axes of the ellipsoid of inertia.  The results to the second order approximation are the same as that obtained in general relativity and consistent with experiments.\cite{1,3} 

The wave equation of the fermion field $\psi$ in Yang-Mills gravity can be derived from the Lagrangian $L_\psi$.  In the geometric-optics limit,\cite{1,2} the fermion wave equation reduces to a Hamilton-Jacobi  type equation,
\be
G^{\mu\nu} (\p_\mu S)(\p_\nu S) -m^2=0, \ \ \  G^{\mu\nu}=\e_{\a\b}J^{\a\mu} J^{\b\nu}.
\ee
This equation of motion for macroscopic objects in flat space-time involves a new effective Riemannian metric tensor $G^{\mu\nu}$, which is actually a function of the $T_4$ gauge field $\phi_{\mu\nu}$ in Yang-Mills gravity.   It  is formally the same as the corresponding equation of motion for macroscopic objects in general relativity and we have named it the `Einstein-Grossmann' equation of motion.\cite{3}  This equation is crucial for Yang-Mills gravity to be consistent with the perihelion shift of the Mercury, the deflection of light by the sun and the equivalence principle.\cite{1,3}
 
A satisfactory theory of gravity should be able to explain why the gravitational force is attractive rather than repulsive.
It is gratifying that, in quantum Yang-Mills gravity, this property is embedded in the coupling between the gravitational tensor field $\phi_{\mu\nu}$ and the matter fermion field $\psi$ at the quantum level in the Lagrangian (6).  Let us consider the  gravitational ($T_4$) tensor field $\phi_{\mu\nu}(x)$ and the electromagnetic potential field $A_\mu(x)$ in the gauge covariant derivative  and its complex conjugate in the fermion Lagrangian (6),
\be
 \p_\mu -ig\phi_\mu^\nu p_\nu -ie A_\mu +.... = \p_\mu +g\phi_\mu^\nu \p_\nu -ie A_\mu +....
\ee
\be
 (\p_\mu -ig\phi_\mu^\nu p_\nu -ie A_\mu +....)^* = \p_\mu +g\phi_\mu^\nu \p_\nu + ie A_\mu +....
\ee
The gauge covariant derivative (9) and its complex conjugate (10) appear respectively in the wave equations of the electron (i.e., particle with charge $e<0$) and the positron (i.e., antiparticle with charge $-e$).  The electric force between two charged particles is due to the exchange of a virtual photon.  In quantum electrodynamics, this can be pictured in the Feynman diagrams with two vertices connected by a photon propagator.  The key properties of the electric force $F_e(e^-,e^-)$ (i.e., between electron and electron) and the force $F_e(e^-, e^+)$ (i.e., between electron and positron) are given by the third terms in (9) and in (10), i.e.,
$$
F_e(e^-,e^-): \ \ (-ie)\times (-ie)= - e^2, \ \ \ \ \   repulsive, 
$$
$$
F_e(e^-, e^+): \ \  (-ie)\times (ie)=+e^2,  \ \ \ \ \ \   attractive,
$$
and the force $F_e(e^+,e^+)$ is the same as $F_e(e^-,e^-)$. Thus, we have experimentally established attractive and repulsive electric forces, which are due to the presence of $i$ in the electromagnetic coupling.  The Yang-Mills gravitational force $F_{YMg}(e^-, e^-)$ (i.e., between electron and electron) and the force $F_{YMg}(e^-, e^+)$ (i.e., between electron and positron) are respectively given by the second terms in (9) and in (10).   Because the gravitation coupling terms in (9) and (10) do not involve $i$, we have only an attractive gravitational force,
$$
F_{YMg}(e^-, e^-) : \ \ \ (g) \times (g)= +g^2, \ \ \ \ \   attractive,  
$$
$$
F_{YMg}(e^-, e^+) : \ \ \  (g) \times (g) =+g^2, \ \ \ \ \   attractive,
$$
and $F_{YMg}(e^+, e^+)$ is the same as $F_{YMg}(e^-, e^-)$.  Note that these qualitative  results for forces $F_e(e^-,e^-)$ and $F_{YMg}(e^-, e^-)$ are independent of the signs of the coupling constants $e$ and $g$.  Furthermore, the gravitational coupling constant $g$ in (9) and (10) has the dimension of length (in natural units), in contrast to all other coupling constants of fields associated with internal gauge groups, so that $g^2$ is related to Newtonian constant $G$ by $g^2=8\pi G$.\cite{7,3}    These important qualitative results revealed through the coupling of the tensor field $\phi_{\mu\nu}$ and the fermion matter field $\psi$ in the Lagrangian (4) appear to indicate  that the space-time translation gauge group of Yang-Mills gravity is just right for gravity.

  It is intriguing that the $T_4$ gauge transformations\cite{1,2} with infinitesimal vector gauge function $\Ld^\mu(x)$ in quantum Yang-Mills gravity turn out to be identical to the Lie derivatives in the coordinate expression.  This indicates an intimate relation between the mathematical theory of Lie derivatives in coordinate expression and the physical gauge field theory with the external space-time translation $T_4$ gauge symmetry.  The vanishing of the Lie derivative of the action $S_{\phi\psi}$ turns out to be the same as the invariance of the action in Yang-Mills gravity under the $T_4$ gauge transformations.\footnote{To prove the $T_4$ gauge invariance of the action of Yang-Mills gravity, Cartan's formula in the theory of Lie derivative facilitates the calculation of the change of the volume element (e.g., $W(x)d^4 x$) under the $T_4$ gauge transformations.  See appendix A in the forthcoming second edition (entitled {\em ``Space-Time, Yang-Mills Gravity and Dynamics of Cosmic Expansion"}, World Scientific, 2019,) of the monograph in ref. 3.}

In the usual macroscopic (i.e., geometric-optics) limit, the wave equations of quantum particles with mass $m$ in Yang-Mills gravity reduces to  the Einstein-Grossmann (EG) equation, $G^{\mu\nu}(x)(\p_\mu S)(\p_\nu S)-m^2=0$. Thus, the apparent curvature of macroscopic spacetime appears to be a manifestation of the flat spacetime translational gauge symmetry for the wave equations of quantum particles in the geometric-optics limit.\cite{2,7}  According to quantum Yang-Mills gravity, macroscopic objects move as if they were in a curved spacetime because their equation of motion involves the `effective metric tensor' $G^{\mu\nu}(x)$, which is actually a function of $T_4$ tensor gauge fields in flat space-time. 

The conventional FLRW model of cosmology\cite{8,9} was based on general relativity to discuss expansion dynamics of the universe.  In quantum Yang-Mills gravity,  the emergence of the effective metric tensor for the motion of macroscopic objects suggests that we can use Yang-Mills gravitational field equations to discuss an alternate dynamics for the expanding universe (i.e., a new HHK model of  particle cosmology\cite{10}). 

\section{Yang-Mills gravity in the supermacro-\\ scopic limit and the Okubo equation}

In the super-macroscopic limit with the cosmological principle of homogeneity and isotropy, the effective spacetime-dependent metric tensor $G^{\mu\nu}(x)$ in the EG equation in Yang-Mills gravity further simplifies to a time-dependent effective metric tensor $G^{\mu\nu}(t)$ with the diagonal form  in an inertial frame $F\equiv F(t,x,y,z), \ c=\hbar=1$.  The $T_4$ gauge field equation and the cosmological principle lead to the solution,\cite{10}
 \be
 G_{\mu\nu}(t)=\left(B^2(t), -A^2(t),-A^2(t),-A^2(t)\right),    
 \ee
 $$
  B=\b t^{1/2}, \ \  A=\a t^{1/2}, \ \ \ \ \ \   c=\hbar=1,
  $$
for matter dominated cosmos, where $\b=3\a^5/2g^2 \rh_o$ and $\a=(8g^6 \om \rh^3_o/9)^{1/12}$. The discussions in  the cosmic HHK model of particle cosmology are based on Yang-Mills gravity within the framework of flat spacetime and the inertial frame $F(t,x,y,z)$.  One advantage is that cosmological predictions of the HHK model are based on well-defined inertial frames of reference, in which space and time coordinates have the usual operational meaning. 

The basic equation of motion of a distant galaxy with mass $m$ can be derived from the principle of least action involving the effective cosmic metric tensor $G_{\mu\nu}(t)$ in (11) and the cosmic action
$
\int (-m ds),
$
where $ds^2=G_{\mu\nu}(t) dx^\mu dx^\nu$.  To derive the cosmic equation of motion, we consider  the covariant  space-time variation $\de x^\nu$  with a fixed initial point and a variable end point and the actual trajectory.\cite{11}  We obtain
$$
\de S_{cos}= \de \int \left(-m \sqrt{G_{\mu\nu}(t) dx^\mu dx^\nu}\right)
$$
\be
= m\int T_\nu \de x^\nu ds + \left[ m G_{\mu\nu}(t) \frac{dx^\mu}{ds}\right] \de x^\nu =p_\nu \de x^\nu. 
\ee
$$
T_\nu = \left[\frac{1}{2}\frac{dx^\mu}{ds} \frac{dx^\ld}{ds}\frac{\p G_{\mu\ld}}{\p x^{\nu}} -\frac{d}{ds}\left(G_{\mu\nu}\frac{dx^\mu}{ds} \right)\right]=0,  
$$
$$
 p_\nu=m G_{\mu\nu}(t) \frac{dx^\mu}{ds},
$$
where the actual trajectory satisfies the equation $T_\nu=0$, and the non-vanishing term in (12) is contributed from the variable endpoint.  Based on (12), we define the generalized four-momentum of an object moving in the super-macroscopic world as the derivative ${\p S_{cos}}/{\p x^\nu}=p_\nu $.\cite{11}  Consequently, we obtain the equation $G^{\mu\nu}(t) p_\mu p_\nu = m^2,$ where $ G^{\mu\nu}(t)G_{\nu\ld}(t)=\de^\mu_\ld$ and $p_\mu$ is given by the last equation in (12).  As usual, substituting $\p S_{cos}/\p x^\mu$ for $p_\mu$ in $G^{\mu\nu}(t) p_\mu p_\nu = m^2$, we find the Okubo equation for the motion  of a distant galaxy with mass $m$:\footnote{This equation of motion (13) for a distant galaxy with mass $m$ derived from the quantum Yang-Mills gravity, together with the cosmological principle, was called the `cosmic Okubo equation' of motion, in memory of his endeavor in particle physics and his `departure ... to the black hole.'} 
\be
G^{\mu\nu}(t) \p_\mu S \p_\nu S  - m^2 =0,   \  \ \ \ \   S\equiv S_{cos}.  
\ee
Thus, we have seen that this cosmic Okubo equation is the generalization of the Einstein-Grossmann equation $G^{\mu\nu}(x)(\p_\mu S)(\p_\nu S)-m^2=0$ in Yang-Mills gravity from macroscopic world to the super-macroscopic world for the motion of distant galaxies and for the expansion of the universe.

\section{Exact recession velocity measured in inertial frames}

Let us explore physical and cosmological implications of the effective metric tensor $G_{\mu\nu}(t)$ and the Okubo equation (13) in the super-macroscopic world ( roughly $\ge$ 300 million lightyears\cite{8}).  Using spherical coordinates, we choose a specific radial direction with specific angles $\th$ and $\phi$ to express (13) in the form,
$$
   B^{-2}(\p_t S)^2 - A^{-2}(\p_r S)^2 - m^2=0. 
$$
 Since $r$ is cyclic in cosmic Okubo equation (13) because $G^{\mu\nu}(t)$ does not involve $r$,\cite{11} we have the conserved `generalized' momentum $p=\p_r S=\p S/\p r$.  As usual, to solve eq. (13), we look for an $S$  in the form\cite{11,3} $S= - f(t) + pr$. We have
\be
f(t) = \int \sqrt{p^2 C_o^2 + m^2 B^2} \ \ dt, \ \ \ \ \ \ \  C_o=B/A=const.
\ee
The trajectory of a distant galaxy is determined by the equation $\p S/\p p = constant.$\cite{11,3} Therefore, we have\footnote{We neglect the constant associated with $r$ in (15) because we are dealing with extremely large distances $r$ in the super macroscopic world and, so far, we have no data to determine its value.}
\be
r-\int \frac{pC_o^2}{\sqrt{p^2C_o^2 + m^2 B^2}}dt = constant.
\ee
Since $B=\b t^{1/2}$, equation (15) leads to
\be
\dot{r}=\frac{dr}{dt}=\frac{pC_o^2}{\sqrt{p^2 C_o^2+ m^2 \b^2 t}}=\frac{C_o}{\sqrt{1+\Om^2 t}}, \ \ \ \  \Om=\frac{m\b}{pC_o},
\ee
\be
r =\frac{2 p C_o^2}{m^2 \b^2} \sqrt{p^2 C_o^2 + m^2 \b^2 t} =\frac{2C_o}{\Om^2}\sqrt{1+\Om^2 t}.
\ee
In the low velocity approximation, i.e., $m>>p$, we have $f(t) \approx (2m \b/3) t^{3/2}$, $r \approx  t^{1/2}({2 p C_o^2}/{m \b}).$ Thus, we obtain the usual Hubble laws (linear in $r$) as the low velocity approximation of the solution to the Okubo equation (13),
\be
\dot{r} = \frac{pC_o^2}{m \b t^{1/2}}=H(t) r, \ \ \ \ \ \  H(t)=\frac{\dot{A}}{A}.
\ee

Suppose we do not take the low velocity approximation.  We have to solve for $\Om^2$ in terms of $r$ and $t$ by using (17),
\be
\Om^2 =\frac{2}{\sqrt{t^2 + r^2/C_o^2} - t},
\ee
and use this result (19) in (16).  Thus, we derive the exact recession velocity $\dot{r}$ in terms of $r H(t) $,
\be
\dot{r}=\frac{rH}{1/2 +\sqrt{1/4 + r^2 H^2/C_o^2}} < C_o,  
\ee
$$
 C_o=\frac{B}{A}= constant.
$$
The upper limit $C_o$ of recession velocity in (20) can be seen as follows:  When the velocity $rH$ is very large, $\dot{r}$ approaches $C_o$, as shown in (20).  We note that such a limiting velocity $C_o$ is the recession velocity at time $t=0$, as one can see in (16) with $t=0$.  Thus, the HHK model predicts that the exact recession velocity is given by (20) with the upper limit,
\be
C_o=\frac{\b}{\a}= (3\om)^{1/3}, \ \ \ \ \ \ \ \   c=\hbar=1,
\ee
where $\om=P/\rh < 1/3$ is the ratio of pressure $P$ to energy density $\rh$ of macroscopic bodies.\cite{10,8}  There is little experimental data for the parameter $\om$.  Presumably, $\om$ is much smaller than 1/3.  For example, suppose $(3\om)$ is $10^{-3}$, the limiting speed $C_o$ is $0.1 (=3 \times 10^7 m/s)$. 

It is natural to interpret (20) as the exact `non-relative' recession velocity as measured in an inertial frame, according to quantum Yang-Mills gravity:  

{\em The recession velocity of a distant galaxy, as measured in an inertial frame, is dictated by the cosmic Okubo equation (13) with the effective metric tensor (11).  It is exactly given by (20) at a given time and has an upper limit $C_o$ in (21) for matter dominated cosmos.  We may interpret $C_o$  as an `effective speed of light' because it is associated with (11) through the vanishing effective metric, $ds^2=G_{\mu\nu}(t)dx^\mu dx^\nu=0$.}

To see the next correction term for $\dot{r}$ in (18), we expand the square-root in (20) and obtain the approximate recession velocity,
\be
   \dot{r} \approx rH(t) \left[1-\frac{r^2 H^2(t) }{C_o^2}\right]. 
\ee
The results (16)-(22) are for the matter dominated cosmos in the HHK model.

For radiation dominated (rd) cosmos, we have different scale factors $A_{rd}(t)$ and $B_{rd}(t)$,\cite{10}
 \be
A_{rd}(t) = \a' t^{2/5}, \ \ \ \ \   \a'=\left(\frac{5^6 g^6 \om \rh_o^3}{2^6\times 6^2}  \right)^{1/15},  
\ee
\be
B_{rd}(t)=\b' t^{2/5}, \ \ \ \ \ \ \  \b'=\frac{24 \a'^6}{25 g^2 \rh_o}, \ \ \ \   g^2=8\pi G,
 \ee
where $G$ is the Newtonian gravitational constant.   As usual, we follow previous steps of calculations with 
$S=-f(t)+pr$, the  recession velocity $\dot{r}_{rd}$ of the radiation dominated (rd) cosmos is 
\be
\dot{r}_{rd}=\frac{pC^{'2}_o}{\sqrt{p^2C^{'2}_o+m^2\b^{'2} t^{4/5}}} < C'_o, \ \ \    C'_o=B_{rd}/A_{rd};
\ee
$$
 \dot{r}_{rd}  \approx \frac{3Hr}{2}\left[1-\frac{9H^2 r^2}{8C^{'2}_{o}}\right] , \ \   H=\frac{\dot{A}_{rd}}{A_{rd}}=\frac{2}{5t}, \ \    m>>p.
  $$
 Based on (23)-(25)  for the radiation dominated cosmos, the HHK model with Yang-Mills gravity predicts the limiting recession speed $C'_o$ to be 
\be
C'_o={\b'}/{\a'}= (6\om)^{1/3},
\ee
which is the maximum Hubble recession speed at time $t=0$ for the radiation dominated cosmos.  
 
 The limiting recession speed (26) for distant galaxies appears to be in harmony  with the 4-dimensional space-time framework of quantum Yang-Mills gravity.   Both limiting speeds (21) and (26) are `effective speeds of light' given by the vanishing effective line element $ds^2=G_{\mu\nu}(t)dx^\mu dx^\nu=0$, where the scale factors are given by (11) for matter dominated universe and by (23)-(24) for radiation dominated universe.
  
\section{Cosmic redshifts measured in inertial frames}
For cosmic redshifts of lights emitted from distant galaxies with non-constant recession velocities (20) or (22) can be treated similar to the Doppler effect or, to be more specific, accelerated Wu-Doppler effects.\cite{12,3}  The exact Doppler shift can be obtained by the transformation of two wave 4-vectors $k'_\mu$ (emitted from a source at rest in a moving inertial frame $F'$) and $k_\mu$ (observed in an inertial frame $F$), which are related by the invariant law $\e^{\mu\nu}k'_\mu k'_\nu =\e^{\mu\nu}k_\mu k_\nu$.  Similarly, the wave 4-vector $k_{e\mu}$ emitted from a distant galaxy is associated with an eikonal equation or the massless Okubo equation $G^{\mu\nu}(t)\p_\mu \psi_e \p_\nu \psi_e=0$ with $\p_\mu \psi_e=k_{e\mu}$.  This is consistent with the (massive) Okubo equation (13) related to distant galaxies.  It is also the generalization of the usual eikonal equation in the macroscopic world to the super-macroscopic world with the effective time-dependent metric tensor $G^{\mu\nu}(t)$ in Yang-Mills gravity. 

The wave 4-vector of light as measured by observers in the inertial frame $F$ satisfies  the usual eikonal equation,   $\e^{\mu\nu}\p_\mu \psi \p_\nu \psi=0$ with $\p_\mu \psi= k_{\mu}$.\cite{11}  The distant galaxy moves with (negative) acceleration and the observer is in an inertial frame $F$.  Based on the principle of limiting continuation of physical laws, the laws of physics in a reference frame $F_1$ with an acceleration ${\bf a}_1$ must reduce to those in a reference $F_2$ with an acceleration ${\bf a}_2$ in the limit where ${\bf a}_1$ approaches ${\bf a}_2$.\cite{3,12}  In the special case  ${\bf a}_2=0$, this principle of limiting continuation of physical laws reduces to the principle of relativity in the zero acculeration limit.  Therefore, it is natural to treat the observed redshift in analogy to the accelerated Wu-Doppler effect with the `covariant' eikonal equation\cite{12,3}
\be
G^{\mu\nu}(t)\p_\mu \psi_e \p_\nu \psi_e= \e^{\mu\nu}\p_\mu \psi \p_\nu \psi,  \ee
$$
 \p_\mu \psi_e= k_{e\mu}, \ \ \  \p_\mu \psi= k_{\mu}.
$$
For simplicity, we choose a specific radial direction in a spherical coordinate with $k_{e\mu}=(k_{e0}, k_e,0,0)$ and
$k_\mu=(k_0, k,0,0)$, similar to (13) so that, for matter dominated cosmos, we have
\be
B^{-2}(t) k_{e0}^2-A^{-2}(t)k^2_{e} = k_0^2 - k^2=0, 
\ee
based on the principle of limiting continuation of physical laws.  The recession velocity of a distant light source is the velocity $V_r$ and, hence, the frequency $k_0$ observed in F decreases.  The relation (28) leads to the transformations for these covariant wave 4-vectors:
\be
\frac{k_{e0}}{B} = \G\left[k_0 +V_r k\right], \ \ \ \
\frac{k_{e}}{A} = \G\left[k + {V_r} k_0\right], 
\ee
$$
  \G=\frac{1}{\sqrt{1-V_r^2}}.
$$
One can verify that the covariant law (28)  is preserved by the transformations (29) with a velocity function $V_r$.  Since a distant galaxy moves with a non-relative velocity (20), by dimensional analysis, it is natural to identify $V_r <1$ with the recession velocity, $V_r=\dot{r}/C_o <1$ in (20). Specifically, we choose $k=+k_0$ for the recession of the radiation source in the radial direction.  

For convenience of explanation, let us introduce an `auxiliary expansion frame' $F_e$ associated with the effective metric tensor $G^{\mu\nu}(t)$ and all distant galaxies are, by definition, at rest\cite{11} in $F_e$. Thus, the  frequency shift for a source at rest in an expansion galaxy is given by
\be
\left[\frac{k_{e0}}{B}\right]_{at\ rest \ in \ F_e}=\frac{k_0(1+V_r)}{\sqrt{1-V_r^2}},     \ \ \ \ \ \   V_r=\frac{\dot{r}(t)}{C_o},
\ee
where $ k_0=[k_0]_{observed \ in \ F}$ and $V_r=\dot{r}(t)/C_o$ is the non-relative recession velocity of the distant galaxy at the moment when the light was emitted.  The frequency shift (30) is formally the same as the accelerated Wu-Doppler shift involving constant-linear-acceleration source.\cite{3}

We stress that the expansion frame $F_e$ is only an `auxiliary frame', which does not have well-defined space and time coordinates over all space.  Therefore, all experiments and observations must be carried out in the inertial frame $F=F(t,x,y,z)$ with operationally defined space and time coordinates.  For experimental test of redshift based on (30), we cannot have data of $k_{e0}$ measured in $F_e$ and, furthermore,  the non-inertial auxiliary frame $F_e$ and the inertial frame $F$ are not equivalent because $F_e$ with the metric tensor $G_{\mu\nu}(t)$ is not an inertial frame in flat space-time.  Therefore, we must express $[k_{e0}/B]_{at \ rest \ in \ F_e}$ in (30) in terms of the same kind of radiation source at rest and observed in the inertial frame $F$.  Here, the situation is again similar to the accelerated Wu-Doppler effects, in which the accelerated frequency shift of a radiation source is at rest in an accelerated frame and observed in an inertial frame.  Thus, we treat $F_e$  as a non-inertial frame and assume weak equivalence of non-inertial frames on the basis of the principle of limiting continuation for physical law.\cite{13,3}  Such a weak equivalence has been supported by Davise-Jennison's two-laser experiments, which involve orbiting laser sources and are observed in inertial laboratory.  According to weak equivalence, we have\cite{13} 
\be
\left[\frac{k_{e0}}{\sqrt{G_{00}}} \right]_{at \ rest \ in \ F_e}=\left[\frac{k_{0}}{ \sqrt{\e_{00}}}  \ \right]_{at \ rest \ in \ F} \ , 
\ee
$$
  \sqrt{G_{00}}=B, 
$$
where  $[ \ ]_{at \ rest \ in \ F}$ should be understood as that  ``the source is at rest in $F$ and its emitted frequencies are measured in the $F$ frame.''  Therefore,  (30) and (31) with $\e_{00}=1$ lead to the following redshift of frequency $k_0=\om$ as measured in the inertial frame $F$,
$$ 
[k_0]_{at \ rest \ in \ F} \equiv k_0(emission)
$$
\be
=k_0(observed)\frac{(1+ V_r)}{\sqrt{1-V^2_r}}.  
\ee
The cosmic redshift z is defined by\cite{8}
\be
\frac{k_0(emission)}{k_0(observed)}=1+z.
\ee
It follows from (32) and (33) that the exact law of the redshift z is related to the recession velocity $V_r $ of a distant galaxy  as follows:
\be
z=\frac{1+V_r}{\sqrt{1-V_r^2}} - 1,  \ \ \ \ \    V_r =\dot{r}/C_o < 1.
\ee
This is the prediction of Yang-Mills gravity for redshift of light emitted by a distant galaxy in terms of its recession velocity.  Note that $V_r=\dot{r}(t)/C_o$ is the non-constant recession velocity of the distant galaxy at the moment when the light was emitted.

 Only for a small recession velocity $V_r << 1$, we have the usual approximate relation\cite{8} with a second-order correction term,
\be
z \approx V_r +\frac{1}{2} V^2_r.
\ee
For matter dominated cosmos, the result (34) enables us to express the recession velocity $\dot{r}/C_o=V_r$ of a distant galaxy in terms of its directly measurable redshift $z$,
\be
V_r=\frac{\dot{r}}{C_o}=\frac{(1+z)^2 -1}{(1+z)^2 +1}=\frac{2z+z^2}{2+2z+z^2} < 1,
\ee
as observed in the inertial frame $F$.  One can also verify that $z=0$ and $z\to \infty$ correspond to $V_r=0$ and $V_r\to 1$ respectively.  If $V_r=\dot{r}/C_o$ is given by 0.5 and 0.8, one has z=0.8 and 2 respectively.

\section{Discussions and summary} 
 Thus, according to the present HHK model\cite{10} of particle cosmology based on Yang-Mills gravity, the Hubble recession velocity $V_r$ of a distant galaxy, as measured in an inertial frame, can only take the values between 0 and $C_o$.  However, the cosmic redshift z can take the values between 0 and infinity in the matter dominated cosmos.  
 
 The Okubo equation of motion for a distant galaxy leads to exact solutions for $r(t)$ and $\dot{r}(t)$ in (17) and (16) for the matter dominated universe with the effective metric tensor in (11).  These solutions sketch a total cosmic history from the beginning to the end of the universe:

 \bigskip
 \noindent
 (i) In the beginning $ t=0$, we have the following features:
 		
 		 	initial mass run away velocity $\dot{r}=C_o$,
		 	
 		 	initial non-vanishing radius $r=2p^2 C_o^{3}/(m^2 \b^{2})\equiv r_o$,
		 
 		 	initial Hubble recession velocity $V_r=\dot{r}/C_o=1,$
		 
 		 	initial cosmic frequency red-shift given by (32), 
 $$
 z=\frac{k_o(emission)}{k_o(observed)}- 1= \infty, \ \ \ \ \  for \ \  V_r=1.
 $$
  
 \bigskip
 \noindent
  (ii) At the end $t\to \infty$, we have the following features:
 
 			final velocity of galaxies $\dot{r} \to 0, $ 
		
 			final radius  $r \to \infty,$ 
		
 			final Hubble recession velocity $V_r \to 0$, 
		
 			final cosmic frequency red-shift $z \to 0, $
 
\bigskip 
\noindent
 where we have used (11), (16), (17) and (34).  These properties will be modified when the quantum nature of Yang-Mills gravity and other new long-range and short-range forces\footnote{For example, the extremely weak linear force of baryon-lepton charges with general Lee-Yang $U_1$ symmetry. Cf. M. Khan, Y. Hao, J.P. Hsu, https://doi.org/10.1051/epjconf/201816804004.} in particle physics are taken into account.   Of course, a more realistic model will not be completely dominated by matter.  It may be dominated by a combination of matter, radiation and some sort of effective `vacuum energy.'  One can imagine that the universe has been doing extremely complicated multi-tasks during its evolution.\footnote{These cosmic multi-tasks are presumably beyond our understanding based on the present-day particle physics and quantum field theory, let alone general relativity, which is incompatible with quantum mechanics and the existence of antiparticles.\cite{6,5}}
 
   Interestingly, it appears that the Okubo equation with $m > 0$ in the HHK model of particle cosmology automatically initiates the mass runaway, which resembles some sort of detonation at time t=0 with a maximum speed, as discussed in (i).

 Within the Big Jets model,\cite{10} which is suggested by the fundamental CPT invariance in particle physics, all the previous results of HHK model hold in our matter `half-universe.'  Moreover, all these results (such as the effective metric tensor (11), Okubo equations of motions (13) for distant galaxies, light rays, and redshifts, etc.) should also hold in the antimatter `half-universe,'\footnote{It is presumably located far away from our matter half-universe.} because CPT invariance implies the maximum symmetry between particles and antiparticles regarding their masses, lifetimes and interactions.\cite{14,10}

 It may be interesting to observe that we assume some local properties of the spacetime in (11) and (13) to derive some properties about the behavior of point-like galaxies at super-macroscopic distances.  This result may not be surprising because the `local' effective metric tensor (11) embodies the super-macroscopic properties of homogeneity and isotropy.  Thus, such a treatment of the physical system of distant galaxies in the super-macroscopic world seems to resemble `Riemann geometry in the large.'\cite{15}  In this sense, Riemann geometry in the large may play a role as the mathematical base for the expanding cosmos in the HHK model of particle cosmology with quantum Yang-Mills gravity. 
 
 In the conventional FLRW model with general relativity, one has the Hamilton-Jacobi equation $g^{\mu\nu}(t)\p_\mu S \p_\nu S -m^2 =0$ with the metric tensor $g^{\mu\nu}(t)=(1,-a^{-2}(t),-a^{-2}(t),-a^{-2}(t)),$ where the scale factor is given by $a(t)=a_o t^{2/3}$ for the matter dominated universe.\cite{8}  Following similar calculations from (13) to (18), one obtains the recession velocity $\dot{r}=(\dot{a}/a)(r/2)$ for low velocity approximation ($m>>p)$.  However, there is no constant upper limit for the recession velocity $\dot{r}$ at large momenta ($p>>m)$.  Since these results are not obtained in an inertial frame, it is difficult to have a satisfactory comparison between these results and those  obtained by Yang-Mills gravity in an inertial frame. 
 
In summary, within Yang-Mills gravity, the effective cosmic metric tensor $G^{\mu\nu}(t)= 
(B^{-2},-A^{-2},-A^{-2},-A^{-2})$ appears to play a more basic and useful role than that of $g^{\mu\nu}(t)=(1,-a^{-2},-a^{-2},-a^{-2})$ in the conventional theory.  The reason is that the Okubo equations $G^{\mu\nu}(t)\p_\mu S \p_\nu S - m^2=0$, with $m > 0,$ and $m=0$ can describe completely recession velocities, $0 \le \dot{r}/C_o <1$, and cosmic redshift z for $0 \leq z < \infty $ without making low velocity approximations.  Furthermore, Yang-Mills gravity suggests new views of the universe:  (A) the linear Hubble law is the low-velocity approximate solution to the Okubo equation for distant galaxies, and (B) the recession velocity has an upper limit, whose numerical value depends on the equation of state, $P=\rh \om$.  Thus, cosmic Okubo equations (13) with $m\ge 0$ are basic equations of motion for the super-macroscopic world according to quantum Yang-Mills gravity.  The predictions of recession velocities and redshifts z in (20), (22), (25), (34) and (10) could be tested experimentally. 
  
  The work was supported in part by Jing Shin Research Fund and Prof. Leung Memorial Fund of the UMassD Foundation.  We would like to thank D. Fine, W. S. Huang and W. T. Yang for useful discussions.

\bigskip  
 
\bibliographystyle{unsrt}

\end{multicols}

\clearpage
\end{document}